\documentclass[twocolumn,prl,aps,showpacs]{revtex4}
\usepackage{graphicx}
\usepackage{epsfig}

\begin{document}

\def\beq{\begin{equation}}
\def\eeq{\end{equation}}

%------------------ --------------------------------------------

\title{Self-Regulation of Solar Coronal Heating Process
via Collisionless Reconnection Condition}

\author{Dmitri A. Uzdensky}
\email{uzdensky@astro.princeton.edu}
\affiliation{Dept.\ of Astrophysical Sciences, Princeton University,
and Center for Magnetic Self-Organization (CMSO), Princeton, NJ 08544}

%\date{\today}
\date{June 29, 2007}

\begin{abstract}
I propose a new paradigm for solar coronal heating viewed as
a self-regulating process keeping the plasma marginally collisionless.
The mechanism is based on the coupling between two effects.
First, coronal density controls the plasma collisionality and
hence the transition between the slow collisional Sweet-Parker
and the fast collisionless reconnection regimes.
In turn, coronal energy release leads to chromospheric evaporation,
increasing the density and thus inhibiting subsequent reconnection
of the newly-reconnected loops. As a result, statistically, the
density fluctuates around some critical level, comparable to that
observed in the corona.
In the long run, coronal heating can be represented by repeating
cycles of fast reconnection events (nano-flares), evaporation episodes,
and long periods of slow magnetic stress build-up and radiative cooling
of the coronal plasma.
\end{abstract}


\maketitle{}

\newpage

%*************************************************************************

This paper is devoted to the problem of solar coronal heating 
(see Ref.~\cite{klimchuk2006} for a recent review), viewed in 
the context of Parker's nano-flare model \cite{parker1988}. 
Since the main heating process in this model is magnetic reconnection, 
I will first summarize the recent progress in reconnection research 
achieved in the past 20 years. 
Even though a complete picture of reconnection is still not available, 
there is now consensus about some of its most important aspects. 
My main goal is to apply this new knowledge to the coronal heating 
problem.

First, I want to emphasize the importance of a realization by Petschek
\cite{petschek1964} that the main bottleneck in the classical Sweet--Parker
\cite{sweet1958,parker1957} reconnection model is the need to have 
a reconnection layer that is both thin enough for the resistivity 
to be important and thick enough for the plasma to be able to flow out. 
Furthermore, Petschek \cite{petschek1964} proposed that this difficulty
can be mitigated if the reconnection region has a certain sub-structure
--- the Petschek configuration, with four shocks attached to a central 
diffusion region. This results in an additional geometric factor leading 
to faster reconnection. 
This idea is especially important for astrophysical systems, including 
the solar corona, irrespective of the actual microphysics inside the layer. 
Indeed, the system size~$L$ is usually much larger than any microscopic 
physical scale~$\delta$, e.g., the ion gyro-radius~$\rho_i$, the ion 
collisionless skin-depth~$d_i \equiv c/\omega_{pi}$, or the Sweet--Parker 
layer thickness~$\delta_{\rm SP}=\sqrt{L\eta/V_A}$. Therefore, a simple 
Sweet--Parker-like analysis would give a reconnection rate $v_{\rm rec}/V_A$ 
scaling as~$\delta/L \ll 1$, and hence would not be rapid enough 
to be of any practical interest. 
Thus, we come to a conclusion that {\it Petschek's mechanism (or a variation 
thereof) is necessary for sufficiently fast large-scale reconnection.}

Recently, however, several numerical and analytical studies 
(e.g., \cite{biskamp1986,scholer1989,ugai1992,ma1996,uk1998,uk2000,gem2001,
erkaev2001,bs2001,kulsrud2001,mlk2005}) 
and laboratory experiments \cite{ji1998} have shown that, in resistive MHD 
with a uniform (and, by inference, Spitzer) resistivity, Petschek's mechanism 
does not work; the slow Sweet--Parker scaling applies instead. In other words,
{\it In the collisional regime, when classical resistive MHD applies, 
one does not get Petschek's fast reconnection.} 

It is then natural to ask now is whether fast reconnection is possible 
in a collisionless plasma where resistive MHD is not valid.
The answer now appears to be ``yes''. 
First, in space and solar physics fast collisionless reconnection events
have been observed for a long time. More recently, it has also been seen 
in laboratory experiments \cite{ji1998,yamada2006}. 
In addition, several theoretical and numerical studies have recently 
indicated that fast Petschek-like reconnection does indeed take place 
in the collisionless regime. 
Moreover, there may even be two physically-distinct mechanisms 
for fast collisionless reconnection:
(1) {\it Hall effect} (e.g., \cite{mandt1994,biskamp1995,ma1996,
shay1998,gem2001,bmw2001,breslau2003,cassak2005});
and 
(2) spatially-localized {\it anomalous resistivity}
(e.g., \cite{ut1977,sh1979,scholer1989,ugai1992,
erkaev2001,kulsrud2001,bs2001,uzd2003,mlk2005}).
At present, it is still not clear which of them operates under what 
conditions and how they interact with each other. However, both of
these mechanisms seem to work and both seem to involve an enhancement 
due to a Petschek-like configuration. Thus, I believe it is safe to 
say that {\it a Petschek-enhanced fast reconnection does indeed happen in 
the collisionless regime.}

To sum up, there are two regimes of magnetic reconnection:
the slow Sweet--Parker reconnection in resistive-MHD with classical 
collisional resistivity, and the fast Petschek-like collisionless 
reconnection.

%-------------------------------------------------------------------

Now, how can one quantify the transition between these two regimes?
First, consider the case with a relatively weak (or zero) guide field, 
$B_{z} \lesssim B_0$, where $B_0$ is the reconnecting field component. 
Then, the condition for fast collisionless reconnection can be formulated 
(e.g., 
\cite{ma1996,kulsrud2001,uzd2003,cassak2005,yamada2006,uzd2006,uzd2007}) 
roughly as 
\beq
\delta_{\rm SP} < d_i \, .
\label{eq-cond-1}
\eeq
(Since the discussion in this paper is very approximate, 
I will consistently ignore all numerical factors of order~1.)

Expressing resistivity that enters via $\delta_{\rm SP}$ in terms 
of the Coulomb-collision electron mean-free path~$\lambda_{e,\rm mfp}$, 
one gets \cite{yamada2006}:
\beq
{{\delta_{\rm SP}}\over{d_i}} \sim 
\biggl({L\over{\lambda_{e,\rm mfp}}}\biggr)^{1/2}\ 
\biggl(\beta\,{m_e\over{m_i}}\biggr)^{1/4} \, ,
\label{eq-yamada-2006}
\eeq
where $\beta$ is the ratio of the plasma thermal pressure ($ 2 n_e T_e$)
at the center of the layer to the reconnecting magnetic field pressure 
($B_0^2/8\pi$) outside of the layer. The condition of force balance across
the layer (in the absence of a strong guide field) requires $\beta\simeq 1$, 
where we neglected the contribution due to the upstream gas pressure. 
Then, the above fast collisionless reconnection condition becomes
\beq
L < L_c \equiv \sqrt{m_i/m_e}\, \lambda_{e,\rm mfp} 
\simeq 40 \,  \lambda_{e,\rm mfp}  \, .
\label{eq-cond-2}
\eeq
Note that, by construction, the mean-free path that enters here is 
due to classical Coulomb collisions. It is given by
$\lambda_{e,\rm mfp} \simeq 7\cdot 10^{7}{\rm cm}\, n_{10}^{-1}\, T_7^2 $, 
where we took $\log\Lambda \simeq 20$ and where~$n_{10}$ and~$T_7$ 
are the central layer density~$n_e$ and temperature~$T_e$ in units 
of $10^{10}\, {\rm cm}^{-3}$ and~$10^7$~K, respectively. 
Substituting this into Eq.~(\ref{eq-cond-2}), we get
\beq
L < L_c(n,T) \simeq 3\cdot 10^{9}{\rm cm}\, n_{10}^{-1}\, T_7^2 \, .
\label{eq-cond-3}
\eeq 

The strong $T_e$-dependence tells us that knowing the temperature is crucial. 
Note that~$n_e$ and~$T_e$ that enter here are those at the center of a 
Sweet--Parker reconnection layer and are not known {\it a priori}. 
Therefore, we would like to cast condition~(\ref{eq-cond-3}) in an 
alternative form that would involve only the ambient plasma parameters, 
such as the far-upstream values of the plasma density, temperature, and 
magnetic field. Now, the cross-layer pressure balance 
($2n_e T_e = B_0^2/8\pi$), valid in the zero-guide-field case, provides 
us with one relationship between the central~$n_e$ and~$T_e$, and so, 
by itself, it is not sufficient. Indeed, this condition only tells us 
that the thermal pressure at the center of the layer needs to be raised 
to a certain level to balance the outside magnetic pressure, but it doesn't 
tell us whether this is achieved by increasing the density or the temperature. 
In order to break this degeneracy, we need to consider also the equation of 
energy conservation. The logic of our model dictates that this analysis be 
done in the collisional Sweet--Parker regime. At the minimum, this analysis
should include ohmic heating and heat advection and an estimate of various
possible energy-loss mechanisms, such as radiation and electron thermal 
conduction. In particular, it can be shown \cite{uzd2007} that: 
{(i)} on the time that a fluid element spends inside the layer 
(the Alfv\'en transit time~$\tau_A=L/V_A$) --- ohmic heating converts to heat 
just enough magnetic energy to raise~$T_e$ to the level required by the 
pressure balance, without the need to increase the density substantially; 
{(ii)} for the solar coronal conditions, radiative losses are negligible 
on the Alfv\'en time-scale; and {(iii)} the energy losses due to parallel 
electron thermal conduction can be neglected provided that the layer is 
collisional, in the sense of Eq.~(\ref{eq-cond-2}); {(iv)} the energy 
losses due to the perpendicular electron thermal conduction are only 
marginally important at best. All these findings lead us to the conclusion 
that, once the condition $L> 40\,\lambda_{\rm mfp}$ is satisfied and hence
the system is in the collisional Sweet--Parker regime, the energy equation
can be regarded basically as a balance between ohmic heating and advection. 
As a result, the plasma density in the center of the layer should remain 
roughly comparable to the ambient coronal density, whereas the temperature 
should increase dramatically. In fact, ohmic heating is enough to raise 
the temperature up to the ``equipartition'' level that depends only on 
the ambient density and upstream magnetic field~$B_0$, and is insensitive 
to the ambient coronal temperature: 
\beq
T_e \sim T_e^{\rm eq} \equiv {{B_0^2/8\pi}\over{2k_B n_e}} 
\simeq 1.4 \cdot 10^7\, {\rm K}\ B_{1.5}^2 \, n_{10}^{-1} \, ,
\label{eq-T_e}
\eeq
where $B_{1.5}\equiv B_0/(30\,{\rm G})$. This estimate applies only as 
long as the resulting value is much higher than the ambient temperature 
(typically of the order of~$2\cdot 10^6$~K for the solar corona), which
means implies that the ambient plasma-$\beta$ with respect to the reconnecting
magnetic field should be~$\lesssim 1$. Using this estimate, the collisionless 
reconnection condition can finally be written as \cite{uzd2006,uzd2007} 
\beq
L < L_c(n,B_0) \sim 6\cdot 10^{9}\,{\rm cm}\ n_{10}^{-3}\, B_{1.5}^4 \, .
\label{eq-cond-4}
\eeq
In this form, the fast-reconnection condition is particularly useful 
because it involves only the global quantities characterizing a given 
reconnecting system: its global length~$L$, the reconnecting component
of the magnetic field~$B_0$, and the ambient plasma density~$n_e$.

%---------------------------------------------------------------

Next, let us consider the strong-guide field case, $B_z\gg B_0$, 
which is in fact more relevant to the problem of solar coronal heating.
Although some of the arguments and results presented above have to be 
modified, conceptually, they remain similar. In particular, the relevant 
collisionless-reconnection scale becomes the ion-acoustic gyro-radius, 
$\rho_s$, calculated with the total magnetic field $B_{\rm tot}\simeq B_z$
(e.g., \cite{egedal2007}). Correspondingly, the collisionless 
reconnection condition becomes \cite{cassak2007}:
\beq
\delta_{\rm SP} < \rho_s \sim
d_i \  \beta_e^{1/2} \  {B_0\over{B_z}} \, ,
\label{eq-cond-guide-1}
\eeq
where, again, $\beta_e$ is based on the central~$n_e$ and~$T_e$ 
and on the upstream reconnecting field component~$B_0$. Once again,
all the quantities entering Eq.~(\ref{eq-cond-guide-1}) are to 
be estimated in the collisional Sweet--Parker regime. To do this, 
first note that, in the strong guide field case one can no longer 
use the cross-layer pressure balance to deduce $\beta_e\sim 1$; 
this is because a relatively slight compression of the guide field 
can always ensure the pressure balance. Moreover, a strong guide field 
effectively makes the plasma incompressible, so that the central~$n_e$ 
is equal to the ambient value. But this still leaves us with the task 
of evaluating the central electron temperature that is needed to determine 
the Spitzer resistivity. It turns out, however, that the above energy-balance 
arguments for a collisional Sweet--Parker layer still apply, at least 
qualitatively \cite{uzd2007}. Therefore, the ``equipartition'' estimate 
for the central temperature, given by Eq.~(\ref{eq-T_e}), should still 
approximately hold; in particular, one should still have $\beta_e\sim 1$. 
Then, one can repeat the procedure outlined above and derive the following 
approximate condition for the transition to fast collisionless reconnection 
in the strong guide case \cite{uzd2007}:
\begin{eqnarray}
L < L_c(n_e,B_0,B_z)  &=&  \sqrt{m_i\over{m_e}}\ \lambda_{e,\rm mfp}\ 
\biggl({B_0\over{B_z}}\biggr)^{2}  \nonumber \\
&\simeq & 6\cdot 10^{9}\,{\rm cm}\, n_{10}^{-3}\, B_{1.5}^4\,  
\biggl({B_0\over{B_z}}\biggr)^{2} \, .
\label{eq-cond-guide-2}
\end{eqnarray}
Thus, the main effect of a strong guide field is to reduce the critical 
global length~$L_c$ by a factor $(B_z/B_0)^2\gg 1$; that is, the 
collisionless reconnection condition becomes harder to satisfy. 
Also, it is interesting to note that, for fixed values of~$n_e$ and~$B_z$, 
$L_c$ becomes very sensitive to the reconnecting field component: 
$L_c\sim B_0^6$.

%*************************************************************************

Let us now discuss the implications of these findings for the solar corona. 
As long as flux emergence and braiding of coronal loops by photospheric 
footpoint motions continue to generate current sheets in the corona, magnetic 
dissipation in these current sheets results in intermittent heating 
\cite{parker1988,rtv1978}. Typical dimensions and field strengths of these 
current sheets are basically determined by the emerging magnetic structures 
and by the footpoint motions. The main focus of this paper, then, is on what 
sets the typical level of the coronal plasma density and how it relates to 
the intermittent nature of energy release in the corona. My main point is 
that {\it coronal heating should be viewed as a self-regulating process 
that keeps the corona marginally collisionless} in the sense of 
Eqs.~(\ref{eq-cond-1})--(\ref{eq-cond-guide-2}) (see \cite{uzd2006,uzd2007}).

As a first example of how this works, consider a coronal current sheet 
with some given fixed~$L$, $B_0$, and~$B_z$. Resolving~(\ref{eq-cond-guide-2}) 
with respect to~$n_e$, we can define a critical density, $n_c$, below which 
reconnection switches from the slow collisional to the fast collisionless 
regime:
\beq
n_c (B_z \gg B_0) 
\sim 2\cdot 10^{10}\, {\rm cm}^{-3}\, B_{1.5}^{4/3}\, L_9^{-1/3} 
\biggl({{B_0}\over{B_z}}\biggr)^{2/3} \, .
\label{eq-n_c-guide}
\eeq
Notice that this value is comparable to the typical densities observed 
in the active solar corona. I argue that this is just not a coincidence. 
Indeed, if at some initial time, the ambient density~$n_e$ is higher 
than~$n_c(L,B_0,B_z)$, then the current layer is collisional and reconnection 
is in the slow mode. Energy dissipation then is weak; hence, the plasma 
gradually cools through radiation and precipitates to the surface. 
The density around the given current sheet drops and, at some point, 
becomes lower than~$n_c$. Then, the system switches to the fast 
collisionless regime and the rate of magnetic dissipation jumps. 
Next, there is an important positive feedback between coronal energy 
release and the density. A large part of the released energy is conducted
to the surface, where it is deposited in a dense plasma. This leads 
to chromospheric evaporation along the post-reconnected magnetic loops, 
filling them with a dense and hot plasma. The density rises and may now 
exceed~$n_c$. This will shut off any further reconnection (and hence heating) 
involving these loops until they again cool down, which occurs on a longer, 
radiative time-scale.
Thus we see that, although highly intermittent and inhomogeneous, 
the corona is working to keep itself roughly at the critical density 
given by Eq.~(\ref{eq-n_c-guide}). In this sense, coronal heating 
is self-regulating \cite{uzd2006}.

As a second example, consider a situation in which the initial density 
is relatively low, so that radiative cooling rate is much slower than 
the footpoint twisting rate. Then one can regard the density as constant
between reconnection events and focus instead on the slow evolution of 
the reconnecting magnetic field, caused by the motion of the footpoints 
(similar to Refs.~\cite{parker1988,cassak2006}). Let us consider, for example, 
a flux tube, anchored on the solar surface at both ends, with a fixed strong 
axial (guide) field~$B_z$ and a fixed volume~$V$. The tube is comprised of 
smaller flux fibrils that are being wrapped around each other by the 
photospheric motions. Over time, this wrapping leads to the formation 
and strengthening of current layers. For simplicity, let us represent 
this process by a linear growth of the reconnecting field component of 
a single current sheet of a fixed length~$L$: $B_0(t) = \gamma t B_z$. 
Here, $\gamma$ parametrizes the rate of twisting. Let us now try to 
follow the evolution of this system. At first, $B_0$ increases steadily 
in time, while the density stays constant. This continues until $B_0$ 
reaches a critical value that depends on $n_e$ according to collisionless 
reconnection condition (for fixed~$L$ and~$B_z$):
\beq
B_c(n_e,L,B_z) \sim 30\ {\rm G}\ L_9^{1/6}\ n_{10}^{1/2} \ B_{z,2}^{1/3} \, .
\label{eq-B_c-guide}
\eeq
where $B_{z,2}\equiv B_z/(100\, {\rm G})$. We assume here that $B_0$ 
always stays well below~$B_z$. As soon as this critical value is reached,
the system switches to the fast reconnection regime and magnetic energy 
$B_0^2/8\pi$ is rapidly dissipated. Importantly, part of this energy is 
not radiated promptly but is transported by parallel thermal conduction 
to the solar surface. This causes an evaporation episode adding new plasma
to the flux tube under consideration. The amount of plasma added is roughly 
proportional to the energy released in a given event, $B_0^2/8\pi \sim
B_c^2(n_e)/8\pi$, which, according to~(\ref{eq-B_c-guide}), is in turn 
proportional to the density in the tube just before reconnection: 
$\delta n_e \sim B_c^2(n_e) \sim n_e$.

Now let us see what happens on a still longer (several hours) time-scale. 
As a consequence of the first fast reconnection event, the current sheet 
is promptly destroyed and~$B_0$ drops back to nearly zero. The field-line 
twisting, however, still continues, and so the process described above 
repeats. This time, however, the plasma density in the tube is higher, 
and hence the critical magnetic field~$B_c$ is larger and takes longer 
time to reach. In particular, taking the twisting rate~$\gamma$ to be 
constant, the time between subsequent reconnection events scales as
$\delta t = \gamma^{-1} B_c(n_e)/B_z \sim n_e^{1/2}$. Therefore, as 
long as the relative increase in density at each step is small, the 
long-term ($t\gg \delta t$) evolution can be effectively described 
by the differential equation:
\beq
{dn_e\over{dt}} \simeq {{\delta n_e(n_e)}\over{\delta t(n_e)}} 
\sim \sqrt{n_e} \, ,
\eeq
and so $n_e(t) \sim t^2$. Correspondingly, the emission measure
of the tube increases as~$t^4$.

This growth will continue until one of the following two effects intervene. 
First, as the density builds up, the critical value of~$B_0$ may become so 
large (a sizable fraction of~$B_z$), that the equilibrium shape of the entire 
loop will be affected. The loop may then undergo the kink instability and
become sigmoidal, which, with further twisting, may result in a large-scale 
eruption with a catastrophic energy release (a large flare).

Alternatively, it may happen that the density just builds up gradually 
to a level large enough for radiative cooling between two subsequent 
reconnection events to become important. Indeed, as the density increases, 
the radiative emission measure increases as~$n_e^2$ and the time~$\delta t$ 
between reconnection events as $n_e^{1/2}$ (see above). Ignoring for simplicity
coronal temperature variations, the amount of thermal energy lost between 
two reconnection events scales as~$n_e^{5/2}$, whereas the amount of thermal 
energy gained after each reconnection event is just proportional to $B_c^2(n_e)
\sim n_e$. At some point, the two will inevitably become comparable. 
Correspondingly, the amount of plasma drained due to the gradual radiative 
cooling will become equal to that pumped back up into the corona by each 
chromospheric evaporation episode. Then, on some long time-scale (but still 
only as long as $\gamma$, $L$, and~$B_z$ remain constant), the evolution of 
the system can be represented by repeated cycles that include fast 
reconnection events, followed by chromospheric evaporation episodes, 
followed by relatively long ($\sim$ 1~hour) periods during which the
free magnetic energy builds up and the plasma gradually cools down.

%-----------------------------------------------------------------

\begin{acknowledgments}

I am grateful to E.~Blackman, P.~Cassak, J.~Goodman, H.~Ji, R.~Kulsrud, 
E.~Parker, M.~Shay, and M.~Yamada for fruitful discussions and encouraging 
remarks.

This work is supported by National Science Foundation Grant 
No.~PHY-0215581 (PFC: Center for Magnetic Self-Organization 
in Laboratory and Astrophysical Plasmas).

\end{acknowledgments}

%
%-----------------------------------------------------------------
%                          REFERENCES
%-----------------------------------------------------------------

{}

\end{document}